# Tuning the band gap of PbCrO$_4$ through high-pressure: Evidence of wide-to-narrow semiconductor transitions

D. Errandonea[1], E. Bandiello[1], A. Segura[1], J.J. Hamlin[2], M.B. Maple[2], P. Rodriguez-Hernandez[3], and A. Muñoz[3]

[1]*Departamento de Física Aplicada-ICMUV, Universitat de València, MALTA Consolider Team, C/Dr. Moliner 50, 46100Burjassot, Spain*

[2]*Department of Physics, University of California, San Diego, La Jolla, CA 92093, USA*

[3]*Departamento de Física Fundamental II, Instituto de Materiales y Nanotecnología, Universidad de La Laguna, MALTA Consolider Team, La Laguna, 38205 Tenerife, Spain*

**Abstract:** The electronic transport properties and optical properties of lead(II) chromate (PbCrO$_4$) have been studied at high pressure by means of resistivity, Hall-effect, and optical-absorption measurements. Band-structure first-principle calculations have been also performed. We found that the low-pressure phase is a direct band-gap semiconductor (Eg = 2.3 eV) that shows a high resistivity. At 3.5 GPa, associated to a structural phase transition, a band-gap collapse takes place, becoming Eg = 1.8 eV. At the same pressure the resistivity suddenly decreases due to an increase of the carrier concentration. In the HP phase, PbCrO$_4$ behaves as an n-type semiconductor, with a donor level probably associated to the formation of oxygen vacancies. At 15 GPa a second phase transition occurs to a phase with Eg = 1.2 eV. In this phase, the resistivity increases as pressure does probably due to the self-compensation of donor levels and the augmentation of the scattering of electrons with ionized impurities. In the three phases the band gap red shifts under compression. At 20 GPa, Eg reaches a value of 0.8 eV, behaving PbCrO$_4$ as a narrow-gap semiconductor.

## 1. Introduction

Recently, interesting behaviors have been reported for chromates when squeezed up to around 10 GPa [1 - 8]. In particular, unusual isomorphic transitions have been detected in $PbCrO_3$ (6 GPa) [2] and $PbCrO_4$ (4 GPa) [7]. In the last compound, a second transition has been reported at 12 GPa [7]. Both transitions have been proposed to be correlated to changes in the electronic structure of $PbCrO_4$. The second transition produces a complete darkening of the sample, suggesting that probably it can be associated with a semiconductor-metal transition [7].

$PbCrO_4$ (the mineral crocoite) crystallizes at ambient pressure in the monoclinic monazite structure (space group $P2_1/n$, Z = 4). The structural arrangement of monazite is based on the nine-fold coordination of the Pb and the four-fold coordination of Cr [9]. Monazite-type oxides form an extended family of compounds. Because of their physical and chemical properties, several applications for these materials are already reported or under development [9 - 13]. These applications include photoconductive dielectric materials, humidity sensing resistors, yellow pigments, and effective solid lubricants. This list has been recently extended including applications as novel cathode materials for lithium-ion batteries [10], as green phosphors for light-emitting diodes [11], and in alternative green technologies [12, 13]. Monazites exist in Nature and because of their incorporation of rare-earth elements they can effectively control the rare-earths distribution in igneous rocks [14, 15]. In addition, they are a common accessory mineral in plutonic and metamorphic rocks. Therefore, the knowledge of the high-pressure (HP) behavior of monazites is very relevant not only for technological applications but also for mineral physics, chemistry, and petrology [16]. Because of the broad interest in monazite-type oxides, they have been lately extensively studied [7, 9, 10 - 19].

Interestingly, as in PbCrO$_4$, a similar isomorphic transition has been found under pressure in monazite-type CeVO$_4$ near 12 GPa [19]. In this case, the transition occurs together with a large drop of the electrical resistivity, which has been associated to a pressure-induced metallization. Therefore, it is clear that the study of the electronic properties of monazite oxides deserves some attention. Here, we report a study of the optical and electronic properties of PbCrO$_4$ under compression. These properties have been explored by electrical transport and optical-absorption measurements. Electronic-structure calculations have been also carried out. The low-pressure and the two high-pressure phases of PbCrO$_4$ have been characterized. All of them show a semiconductor behavior, but a substantial narrowing of the band gap is detected upon compression. The technical details of the experiments and calculations are described in Sec. 2. The results are presented and discussed in Sec. 3. Finally, we summarize our conclusions in Sec. 4.

## 2. Experimental methods and computational details

Optical-absorption measurements were carried out in a diamond-anvil cell (DAC) up to 21 GPa. For these experiments, we used 20-µm-thin platelets cleaved from natural crystals of PbCrO$_4$. Electron microprobe analysis showed that the only impurity detectable in the crystals was Fe (0.06 %). Measurements in the visible-near-infrared range were done in an optical setup, which consisted of a tungsten lamp, fluorite lenses, reflecting optics objectives, and several Ocean Optics USB4000 spectrometers [20]. As we will show in Sec. 3, the absorption edge of PbCrO$_4$ moves considerably with pressure demanding the use of different spectrometers to optimize data acquisition. In the experiments a 16:3:1 methanol-ethanol-water mixture was used as a pressure-transmitting medium [21, 22] and ruby chips evenly distributed in the pressure chamber were used to measure pressure by the ruby fluorescence method [23, 24].

Resistivity measurements at room temperature were carried out up to 35 GPa using a "designer DAC" [25]. In these experiments, the pressure was measured by the ruby fluorescence technique [23, 24]. Experiments were performed using high-purity polycrystalline samples obtained from Alfa-Aesar and no pressure medium. The resistance of $PbCrO_4$ was measured employing a standard DC four-probe technique with the four contacts separated by identical distance. The electrical resistivity ($\rho$) was estimated from the measured resistance (R) and the sample thickness (t) using equation $\rho = \pi t R / \ln 2$ [26].

Hall-effect and resistivity measurements under pressure up to 11 GPa were made with a 150 ton hydraulic press and steel-belted Bridgman tungsten carbide (WC) anvils [27, 28]. Measurements were carried out in a Van der Pauw configuration. The samples were contained using two annealed pyrophyllite gaskets. Hexagonal boron nitride was used as pressure-transmitting medium and to electrically isolate the sample from the WC anvils. Samples were compact pellets made from high-purity $PbCrO_4$ powder obtained from Sigma-Aldrich. The pressure applied to the sample was calibrated against the transition pressures of calibrants [29].

*Ab initio* band-structure calculations were performed within the framework of density-functional theory, using the plane-wave pseudo-potential method as implemented in the VASP package [30, 31]. The atomic species were described with projected-augmented wave (PAW) pseudo-potentials [32, 33]. The exchange-correlation energy was taken in the generalized-gradient approximation (GGA) with the PBEsol prescription [34]. The plane-wave basis set employed extended up to an energy cutoff of 520 eV in order to have an accurate total-energy convergence of 1-2 meV per formula unit. We use a dense Monkhorst-Pack [35] grid appropriate to the structures considered to sample the Brillouin Zone (BZ), ensuring a high total-energy convergence. At each selected volume, the atomic positions and the unit-cell parameters were fully optimized. For the relaxed structure at the optimized configuration the atomic forces are smaller than 0.006 eV/Å, and the deviation of the stress tensor from a diagonal hydrostatic form is less than 0.1 GPa [36]. It is important to note that for each volume we obtain the total energy as function of volume, but also energy derivatives, like pressure as function of volume, are obtained from the stress tensor. Band-structure calculations have been carried out up to 12 GPa. They described the two monazite-type polymorphs that exist in this pressure range [7]. Calculations at higher pressures were not performed because the crystal structure of $PbCrO_4$ beyond 12 GPa remains yet undetermined [7].

Due to the heavy mass of the Pb atoms, we also performed, at selected pressures, calculations of the electronic band structure including the spin-orbit (SO) coupling [37]. This greatly increases the computational time required for the simulations. However, from the results we concluded that the SO coupling effects do not appreciably affect the results obtained without including SO effects. Therefore, to evaluate the effect of pressure on the band structure of $PbCrO_4$ calculations were performed neglecting the SO interaction.

**3. Results and discussion**

As described in Ref. [7], at ambient pressure we found that in $PbCrO_4$ the optical-absorption spectrum shows a steep absorption edge characteristic of a direct band gap. The determined value of the band-gap energy (Eg) at ambient pressure is 2.3 eV. We followed the evolution with pressure of the absorption spectrum and determined the pressure dependence of Eg. At all pressures, the absorption spectra suggest that $PbCrO_4$ remains being a direct gap semiconductor. The pressure evolution of Eg is shown in Fig. 1. We found that upon compression the absorption edge gradually red-shifts up to a pressure close to 3 GPa. At 3.5 GPa, an abrupt change is detected, which produces a darkening of the red color of PbCrO$_4$. This fact is caused by the occurrence of a 0.3 eV band-gap collapse (see Fig. 1). As we will explain when discussing calculations, this bang-gap change can be correlated with the isomorphic transition previously detected in PbCrO$_4$ by means of x-ray diffraction experiments (to a HP monazite-type structure) [7]. A similar transition has been also found in other monazites [19, 38, 39]. Beyond 3.5 GPa the pressure evolution of the absorption edge is also toward low energy. When pressure exceeds 12 GPa, the formation of defects in the PbCrO$_4$ crystal is detected. Beyond 14 GPa the crystal suddenly becomes black. This fact is in agreement with the occurrence of a second pressure-driven transition to a phase with an undetermined crystal structure [7]. The blackening of PbCrO$_4$ is caused by the reduction of Eg to values smaller than 1.2 eV, without any evidence of metallization. From 15 to 21 GPa, Eg gradually decreases with pressure from 1.2 to 0.8 eV. In the three phases of PbCrO$_4$, the pressure dependence of Eg can be well described by a linear function. The pressure coefficients, dEg/dP, are equal to -46 meV/GPa, -42 meV/GPa, and -60 meV/GPa, for monazite and the two subsequent HP phases, respectively. Changes induced by pressure in the absorption spectrum are reversible with a hysteresis that can reach up to 2 GPa. It is interesting to note that in the 21 GPa range covered by the experiments PbCrO$_4$ changes from having a band-gap energy similar to that of CdS at ambient pressure to having Eg similar to GaAs in the first HP phase and then having a narrow band gap like InN in the second HP phase.

We will discuss now the results of band-structure calculations in order to explain the behavior observed for Eg below 15 GPa. For the second HP phase such calculations were not carried out because the crystal structure of this phase has not been determined yet. For the calculated pressures, we found that theory agree well with experiments [7], providing a good description of the crystal structure at all the studied pressures. The calculated band structure and density of states (DOS) are shown in Figs. 2 and 3. At ambient pressure, calculations indicate that $PbCrO_4$ is an indirect band-gap material with the maximum of the valence band in between the Y and A points of the BZ and the minimum of the conduction band at the Y point. Calculations also established that the Y-Y direct band gap is 0.05eV higher than the indirect band gap. This fact is in apparent contradiction with experiments ($PbCrO_4$ behaves as a direct gap material). However the energy difference between the direct and indirect gap is small enough to make the indirect band gap undetectable by experiments because of the much higher absorption coefficient of the direct gap. The same apparent contrast between calculations and experiments has been observed in monazite-type $LaVO_4$ [40]. According to calculations, the value of Eg at ambient pressure is 1.3 eV. Therefore, Eg is underestimated by the calculations by 1 eV, as typically occurs with density-functional theory calculations [37]. However, these calculations provide an accurate description of the pressure evolution of Eg [41].

In Fig. 3 it can be seen that at ambient pressure the main contribution to the bottom of the conduction band is coming from the hybridization between Cr 3*d* and O 2*p* orbitals (anti-bonding states). Pb 6*p* states have also a minimal contribution. On the other hand, the upper portion of the valence band is composed basically by the anti-bonding combination of the Pb 6*s* and O 2*p* states. According to our calculations, in the low-pressure phase, under compression the top of the valence band shifts toward high energies faster than the bottom of the conduction band. This is a consequence of the fact that the application of pressure decreases the bond distances (in special Pb-O distances [7]), thus increasing the orbital overlap. As a consequence of it, the separation between Pb bonding and anti-bonding states is enlarged. This fact enhances the displacement towards higher energies of the top of the valence band, but reduces the displacement of the bottom of the conduction band. This behavior is qualitatively similar to what is observed in PbWO$_4$ [42]. It causes a reduction of the energy difference between the bottom of the conduction band and the top of the valence band, inducing the Eg reduction we observed in the experiments up to 3 GPa. Regarding, the band-gap collapse observed at 3.5 GPa, in agreement with experiments, calculations show that the isomorphic transition observed at the same pressure [7] implies a 0.2 eV reduction of Eg. Although the isomorphic structural transition does not affect the global symmetry of the crystal, the crystal structure is highly distorted, affecting Cr-O and Pb-O bond angles and distances [7]. These changes of the crystalline structure are reflected in the electronic structure of PbCrO$_4$, producing the small collapse of Eg that we observed in the experiments. The changes in the electronic structure can be seen by comparing the band structure and DOS at ambient pressure and 8 GPa. In particular, after the transition the conduction band becomes broader. In addition, as observed in the experiments, calculations also found that the HP monazite-type phase has a nearly direct band gap at the Y point of the BZ. Regarding the pressure evolution of Eg in this phase, it is still as a first approximation controlled by the movement towards high energies of Pb 6*s* states. Because of that, after the transition, dEg/dP is similar to that of the low-pressure monazite-type phase.

Fig. 4 shows the resistivity as a function of pressure for PbCrO$_4$ for DAC experiments. These measurements were limited to pressures larger than 8 GPa due to the high resistivity (> 100 MΩ) of the sample below this pressure, which compromise the accuracy of the transport measurements. From 8 to 12 GPa a two-orders of magnitude decrease of the resistivity is observed. A similar behavior is found in the resistivity of related monazite-type CeVO$_4$ [19]. The abrupt decrease of ρ is followed by a plateau between 11 and 14 GPa and a gradual increase of ρ from 14 to 35 GPa. The three pressure regions where the resistivity shows qualitatively different behaviors are consistent with the stability of the three phases of PbCrO$_4$ determined from x-ray diffraction and Raman spectroscopy [7] and with the three pressure regions determined from the band-gap behavior (see Fig. 1). An interesting fact to remark here is that PbCrO$_4$ up to 35 GPa does not show resistivity values compatible with a pressure-induced metallization. On the other hand, the gradual evolution of $\rho$ beyond 14 GPa does not give any hint on the possible occurrence of any additional phase transition in PbCrO$_4$.

Fig. 5 shows the resistivity ($\rho$), carrier concentration (n), and mobility ($\mu$) measured in PbCrO$_4$ as a function of pressure from 7 to 11 GPa (i.e. when the HP monazite-type phase is stable). Measurements at lower pressure cannot be performed accurately due to the high resistivity of PbCrO$_4$. They were not possible at higher pressure due to limitations of the experimental set up [27]. Within the explored pressure range PbCrO$_4$ behaves as an n-type semiconductor. This fact agrees with the semiconductor behavior found in monazite-type LaCrO$_4$ [43]. The carrier concentration monotonically increases with pressure and the mobility gradually decreases up to 9.5 GPa, decreasing considerably more beyond this pressure. The reduction of $\rho$ under compression is then mainly due to the increase of n.

Before discussing the physical reasons of the results described above, we would like to remark the fact that in the three experiments here described different pressure-transmitting medium were used (16:3:1 methanol-ethanol-water, hexagonal BN, and no pressure medium). In spite of it, the experiments corroborate each other, indicating that the changes observed in the electronic properties are independent of the pressure gradient in the sample.

We will now pass to discuss the possible origin of the carrier concentration observed in PbCrO$_4$. The smallest value observed for it is four orders of magnitude larger than the intrinsic concentration for such band-gap range at ambient pressure. Therefore, it is reasonable to assume that the HP phase of PbCrO$_4$ behaves like an extrinsic semiconductor with donor levels. It is know that oxygen vacancy formation is typical in monazite-type oxides [44, 45]. Basically, oxygen migrates from its lattice position to an interstitial site. It creates a vacancy defect at its original site and an interstitial defect at its new location (Frenkel defect). The vacancy could probably behave as a deep donor. This phenomenon has been observed in related oxides [46] where doubly ionized oxygen vacancies and interstitials lead to an n-type semiconductor behavior. In the case of PbCrO$_4$, we think that, as in other compounds [47], the pressure-induced (isomorphic) transition could lead to the creation of a large concentration of donor centers that makes the material to behave as an extrinsic n-type semiconductor. In particular, if one defect is introduced per unit-cell, the concentration of donors can be estimated to have a maximum value of 3 10$^{21}$ cm$^{-3}$, more than enough for creating the measured carrier concentration. Under this hypothesis, let us discuss if the increase found for the carrier concentration can be explained by a decrease of the activation energy of the donors.

Given the existence of Frenkel pairs the material would necessarily contain some compensating acceptors. For a non-degenerate compensated n-type semiconductor, in the low-temperature limit: $n = N_C \dfrac{N_D - N_A}{2 N_A} e^{\frac{-\Delta E_D}{kT}}$, where N$_C$ is the effective density of states of the conduction band, N$_D$ (N$_A$) is the donor (acceptor) concentration, T is the absolute temperature, k is the Boltzmann constant, and $\Delta E_D$ = E$_C$ - E$_D$ is the ionization energy of the donor level (being E$_C$ the energy minimum of the conduction band and E$_D$ the energy of the donor level). From this expression, if the change in the effective density of states is neglected, from the exponential increase of n under pressure we can easily deduce that

$$\frac{\partial \ln n}{\partial P} = -\frac{1}{kT}\frac{\partial (E_C - E_D)}{\partial P}.$$

From the logarithmic slope of the pressure dependence of n we can determine the pressure coefficient of the donor level ionization energy, that turns out to be about 23meV/GPa, which leads to a decrease of some 115 meV in the donor ionization energy in the range from 6 to 11 GPa. Given that the band gap in the first high-pressure phase decreases with a pressure coefficient of -42 meV/GPa, the relative decrease of the band gap in the same pressure interval would be about 13%. The k.p model [37] would predict an overall relative reduction of the electron effective mass by around 13%, leading to the same decrease in the ionization energy of a shallow donor that would not be enough to give account of the observed decrease of the ionization energy. A deep-to-shallow transformation in the nature of the donor level seems more likely as the origin of the large increase of n. This kind of effect is not uncommon in semiconductors and is related to changes induced by pressure in the relative position of different conduction-band minima [25]. The same ionized impurity center gives rise to impurity levels with different ionization energies whose relative position in the band gap depends on the electron effective mass in the corresponding conduction-band minimum. The above described shallow-to-deep donor transformation will drive an increase of the number of ionized donor impurities under pressure leading to the detected increase in n (and the consequent decrease in ρ).

The picture described above, which should be confirmed by combined HP-HT transport measurement [48], provides not only a plausible explanation to the observed behavior of the carrier concentration, but also is consistent with the changes observed in µ. Both facts indicate that it is reasonable model. In particular, the decrease observed for the mobility is consistent with the increase of the ionized impurities concentration leading to an enhancement of electron scattering by ionized donors [49]. The change of slope at about 9.5 GPa would correspond to the change of the dominant scattering mechanism. Below 9.5 GPa phonon scattering is dominant and µ slowly decreases due to the increase of the impurity scattering contribution. Between 9.5 GPa and 11 GPa the mobility decreases by a factor 3, while n increases by the same factor. This indicates that ionized impurity scattering becomes dominant and µ decreases as the inverse of the impurity concentration.

Regarding the increase observed in ρ at the second phase transition (near 15 GPa), it cannot be related to changes in the band structure and the activation energy of donors. Note that $E_g$ is further reduced after the second transition reaching values smaller than 1 eV. A reasonable argument to explain the upturn of ρ is that the second phase transition induced by pressure creates a highly defective crystal. This assumption is consistent with the fact that black defects in the crystal can be visually detected in the pressure range where the onset of the increase of ρ is found (as described above). Indeed, the density of defects rises upon compression and the sample becomes polycrystalline. Carrier trapping and scattering at the grain frontiers can easily induce a decrease in µ. It can also produce a partial compensation of the semiconductor reducing consequently n. Both facts will favor the increase of the resistivity shown in Fig. 4. This hypothesis is consistent with the fact that defect creation has been found to be the cause of pressure-induced resistivity increases in materials with a small band gap [27, 47] (like the second HP phase of $PbCrO_4$).

## 4. Conclusions

The transport and optical properties of PbCrO$_4$ have been explored under compression by means of resistivity, Hall-effect, and optical-absorption measurements as well as band-structure calculations. PbCrO$_4$, a direct band-gap semiconductor with Eg = 2.3 eV, undergoes changes in its electronic properties associated to pressure-induced structural transitions. In particular, at 3.5 GPa a collapse of Eg is detected, which correlates well with a decrease of resistivity and with an isomorphic structural phase transition. Up to the second transition, PbCrO$_4$ behaves as an n-type semiconductor with a donor level. The origin of this level is proposed to be associated to the formation of oxygen vacancies. Upon compression the donors apparently transform from deep to shallow levels producing an increase of the carrier concentration and the observed decrease of ρ. After the second transition (~ 15 GPa) Eg becomes smaller than 1.2 eV. In this phase the resistivity increases as pressure does probably due to the self-compensation of the donor levels and the increase of electron scattering by ionized donors. In the three phases the band gap red shifts under compression. At 20 GPa, Eg is approximately 0.8 eV, so PbCrO$_4$ behaves as a narrow-gap semiconductor. Such a large change of the electronic band gap in a reduced pressure range is associated to the presence of anti-bonding Pb states in the top of the valence band. In spite of the large decrease of Eg induced by pressure, no evidence of metallization is detected in PbCrO$_4$ up to 35 GPa. The conclusions extracted from this study are quite useful for the application of PbCrO$_4$ to selective hydrogen combustion, which open up opportunities for developing energy-efficient oxidative dehydrogenation routes to commercial important olefins [50]. We expect the interesting reported results will probably trigger future studies.

**Acknowledgements**

This study was supported by the Spanish MINECO, Grants No: MAT2010-21270-C04-01/03 and CSD2007-00045. The computing time provided by Red Española de Supercomputación (RES) and MALTA-Cluster is also acknowledged. Designer anvil experiments at UCSD were supported by DOE/NNSA Grant No. DE-FG52-09NA29459. We thank S. T. Weir and Y. K. Vohra for providing the designer diamond anvil.

**Figure Captions**

**Fig. 1:** Pressure dependence of Eg. Symbols: experimental results corresponding to three independent experiments (triangles, circles, and squares). Solid symbols: upstroke. Empty symbols: downstroke. Lines are linear fits to the data.

**Fig. 2:** Calculated band structure of monazite-type PbCrO4 and ambient pressure (top) and 8 GPa (bottom).

**Fig. 3:** (color online) Calculated partial and total DOS of $PbCrO_4$. (top) low-pressure monazite at ambient pressure (bottom) high-pressure monazite at 8 GPa.

**Fig. 4:** Resistivity as a function of pressure measured in the DAC.

**Fig. 5:** Resistivy, carrier concentration, and mobility as a function of pressure measured using large Bridgman anvils.

**Fig. 1**

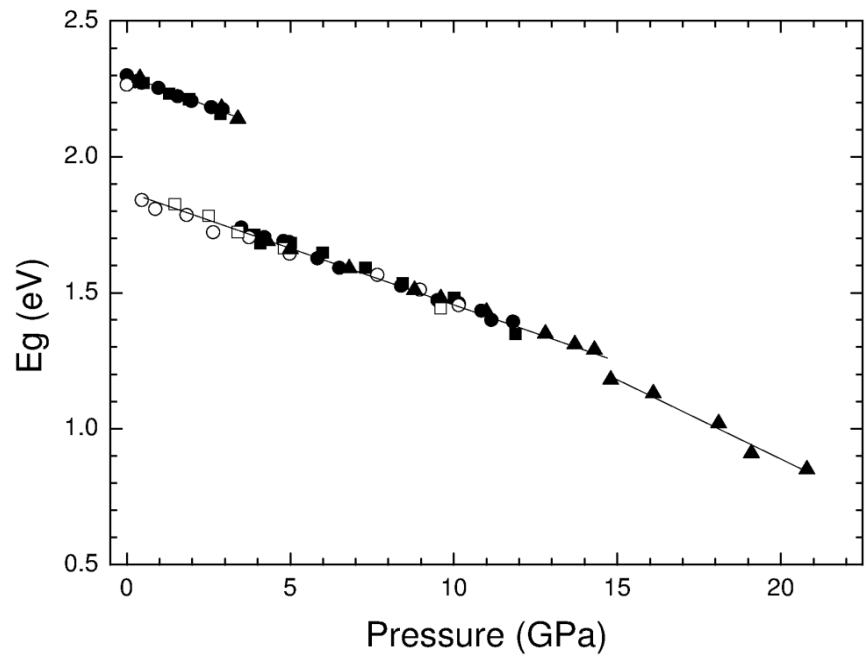

**Fig. 2**

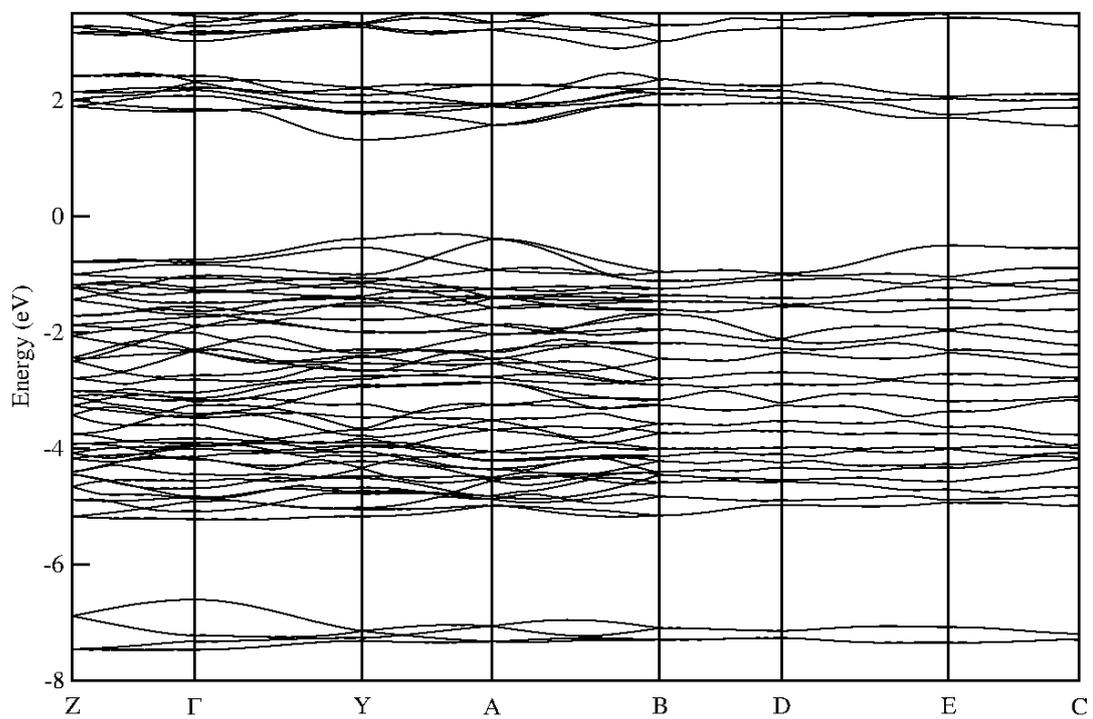

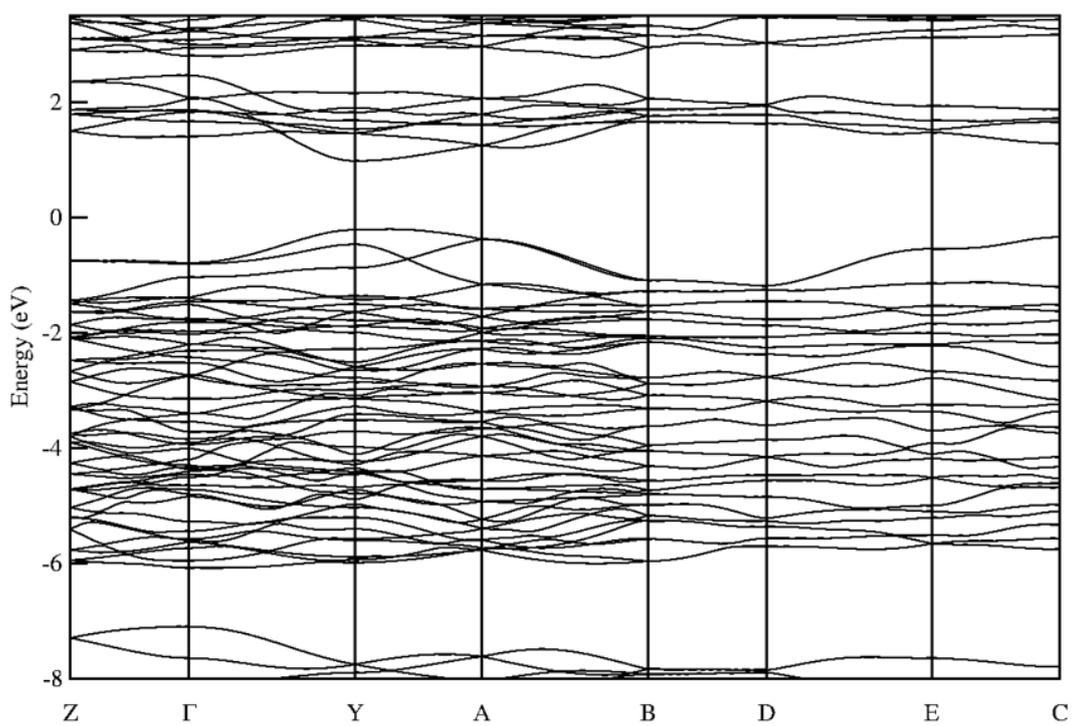

**Fig. 3**

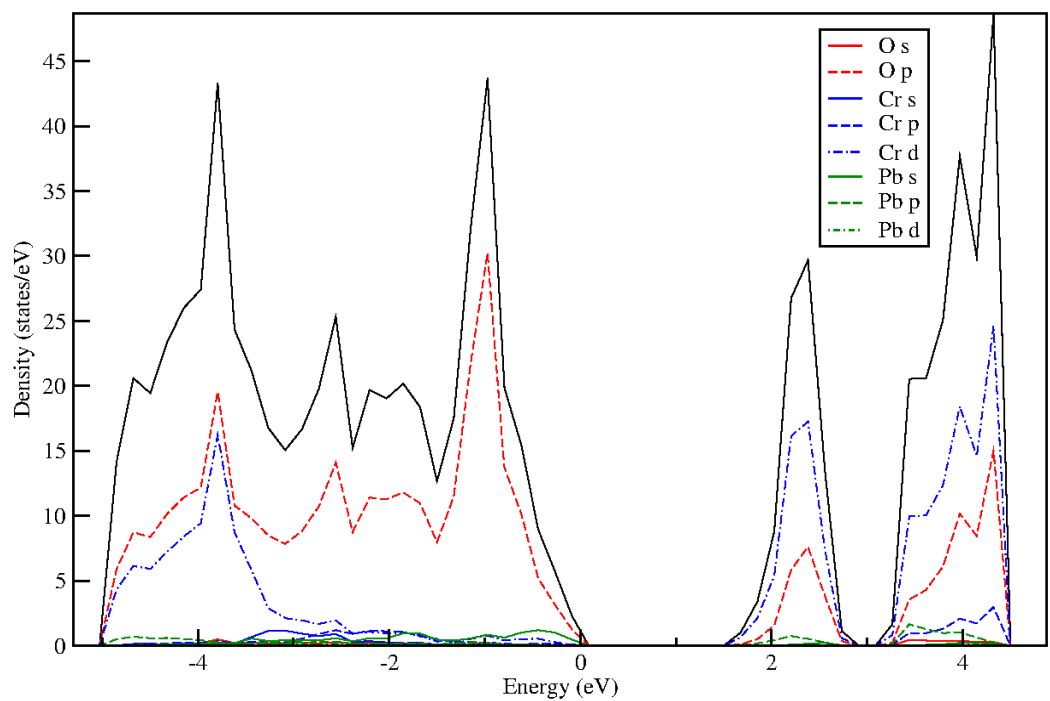

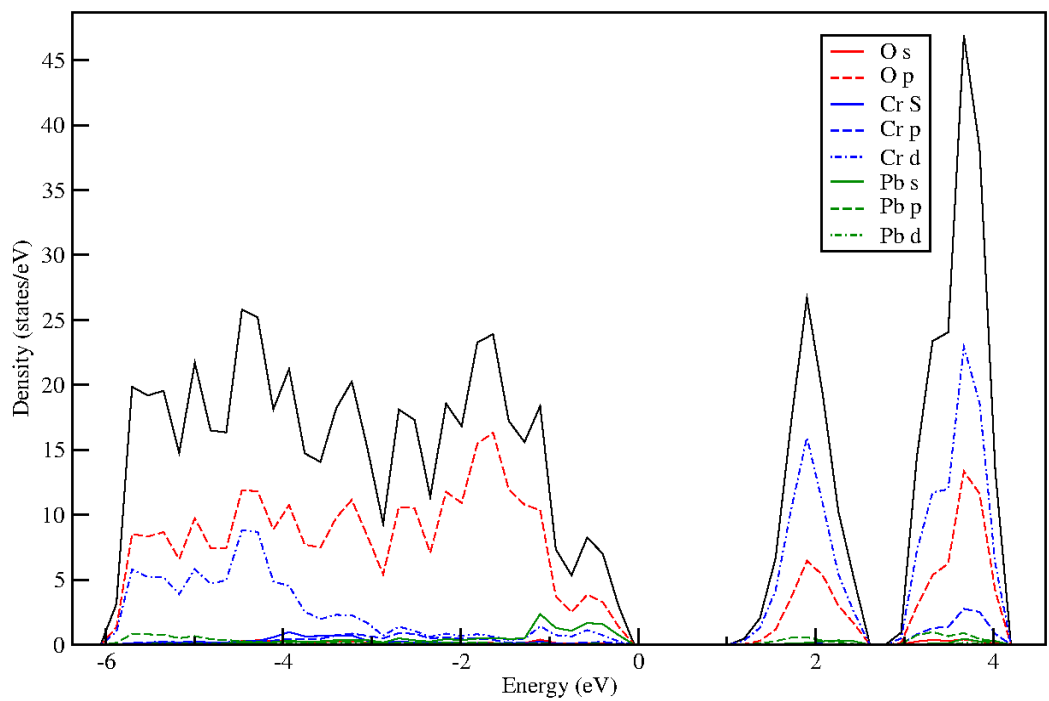

**Fig. 4**

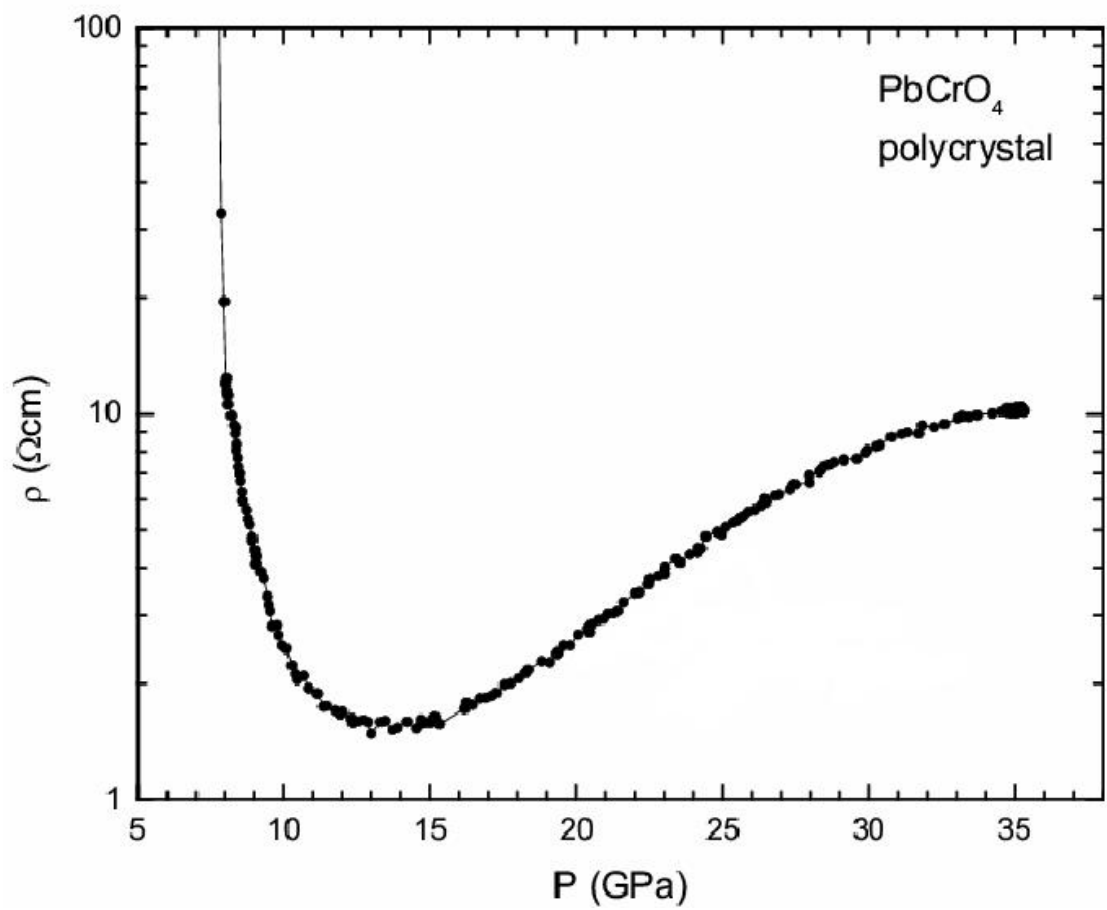

**Fig. 5**

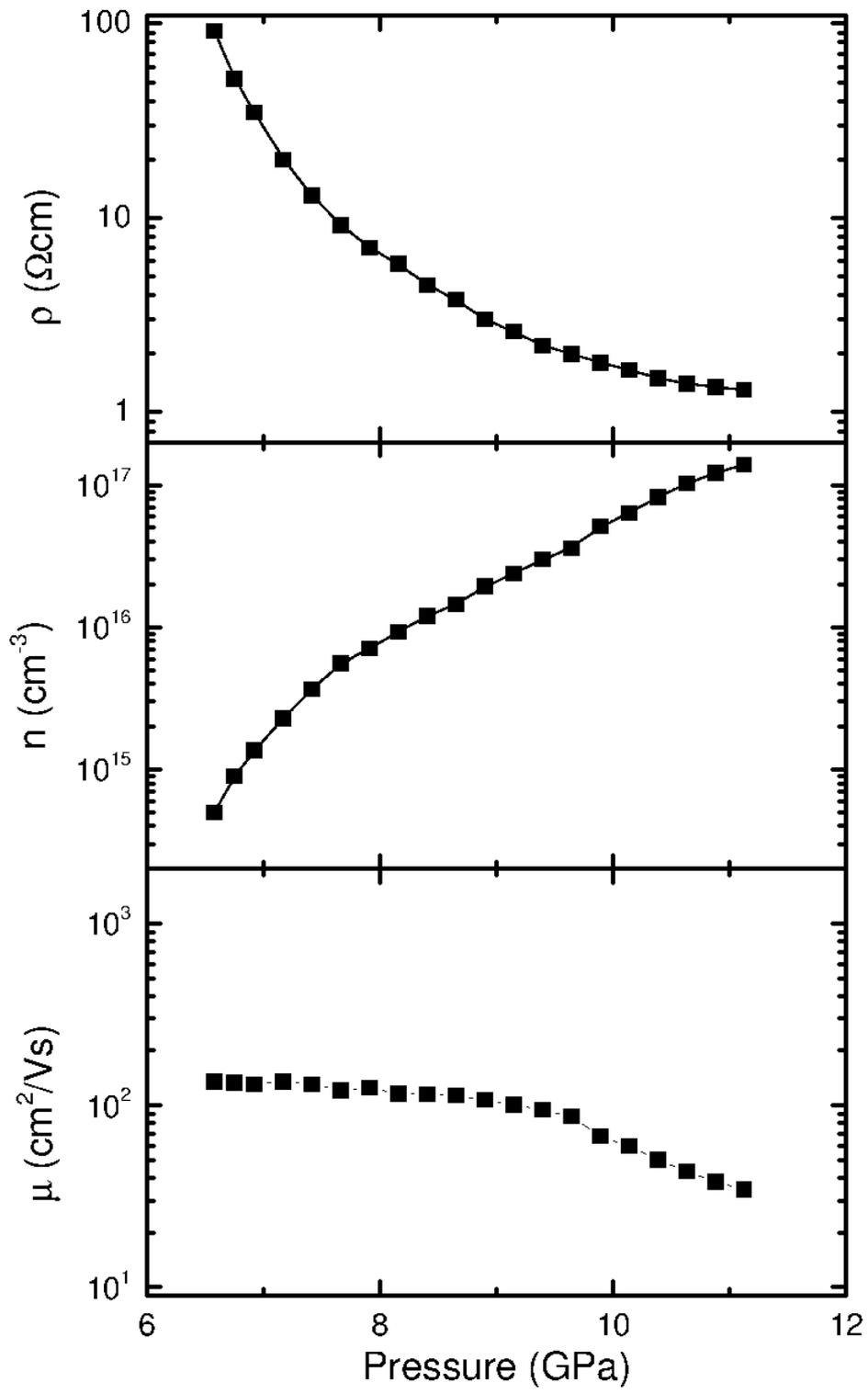